\documentclass[epj]{svjour}

\usepackage{graphicx}
\usepackage{fancyhdr}
\usepackage{axodraw}
\def\makeheadbox{{
 }}
\def\mytitle{My title} 
\def\myauthors{My name}  
\def\mytype{My type of session}
\def\mysession{My session}

\def\mytitle{Gravitino DM with Stop NLSP} 
\def\myauthors{Yudi Santoso}    
\def\mytype{Contributed Talk}    
\def\mysession{Cosmology and Astrophysics}

\def\mstop{m_{\tilde t_1}}
\def\gappeq{\mathrel{\rlap {\raise.5ex\hbox{$>$}}
{\lower.5ex\hbox{$\sim$}}}}
\def\lappeq{\mathrel{\rlap{\raise.5ex\hbox{$<$}}
{\lower.5ex\hbox{$\sim$}}}}

\begin{document}
\title{Gravitino Dark Matter with Stop as the NLSP}
\author{Yudi Santoso 
\thanks{This talk is based on works in collaboration with J.L.~Diaz-Cruz,
John~Ellis and Keith~A.~Olive~\cite{stopNSP}. I would like to thank Xerxes Tata for useful conversation.}
\thanks{\emph{Email:} santoso@uvic.ca, \emph{current address}: IPPP, Department of Physics, University of Durham, Durham DH1 3LE, UK.}%
}                     
%
%
\institute{Department of Physics and Astronomy, University of Victoria,
Victoria, BC V8P 1A1, Canada
}
%
\date{}
\abstract{
Gravitino is one feasible candidate for the dark matter in supergravity models.
With its couplings being suppressed by the Plank mass, gravitino interacts very weakly with other particles, making its direct detection, or production at colliders practically impossible. The signatures of this scenario mainly arise from the next lightest supersymmetric
particle (NLSP) which is metastable due to the very weak coupling. There are many possibilities for the NLSP, and here I will review the feasibility of stop particle as
the NLSP and discuss its phenomenology.    
} 
\maketitle
\section{Introduction}
\label{intro}
Even with many astrophysical evidence pointing to the existence of dark matter,
we have not yet known what the dark matter is. On the other hand, studies in particle physics suggest that it could be a new undiscovered particle from beyond the standard model that is massive, neutral and stable (or very stable with lifetime longer than the age of the Universe). In addition, because of the so far nil results of direct detection searches, the dark matter particle can only interact weakly (or very weakly) with ordinary particles.  

In supersymmetric models with $R$-parity conservation the lightest supersymmetric
particle (LSP) is stable. If the LSP is neutral it could be a candidate for dark matter. Indeed in the minimal supersymmetric standard model (MSSM),
for example, we have neutralino and sneutrino as candidates for dark matter. 
There have been lots of studies on these possibilities, and it would be impossible to list all of the papers on these subjects here. While `left handed' sneutrino dark matter is already excluded, neutralino is still a feasible and appealing candidate for dark matter.    
Including gravity in supergravity models, 
gravitino is another neutral particle that has been shown as a suitable
candidate for the dark matter, see e.g.~\cite{feng,gdm}.
The strength of gravitino interaction with other particles is suppressed by the
Planck scale and  as a consequence the next lightest suppersymmetric particle
(NLSP) would be metastable, assuming that $R$-parity is conserved. The
phenomenology of this scenario depends much on what the NLSP is. Depending on
the supersymmetric model, it could be any of the supersymmetric particles
beside the gravitino, each with its own phenomenology. 
Here we would like to discuss the scenario with the lighter stop particle as the NLSP~\cite{stopNSP}. We will show that it has distinct phenomenology, and discuss its impact on cosmology and supersymmetry searches at colliders.

\section{Properties of Stop NLSP}

Let us start by looking at the stop mass matrix. In the left-right basis the mass matrix has the following form
\begin{equation}
\widetilde{M}^2_{\tilde{t}} = \left(
         \begin{array}{ll}
          M_{LL}^2         &  M_{LR}^2 \\ 
          M_{LR}^{2\,\dag}   &  M_{RR}^2
         \end{array}
         \right)
\end{equation}
where the components are 
\begin{eqnarray}
M_{LL}^2 &=& M_{\tilde{t}_L}^2
+m_t^2+\frac{1}{6}\cos2\beta \,(4m_W^2-m_Z^2) \nonumber \\
M_{RR}^2 &=& M_{\tilde{t}_R}^2+m_t^2
+\frac{2}{3}\cos2\beta\sin^2\theta_W\, m_Z^2 \nonumber \\
M_{LR}^2 &=& -m_t (A_t + \mu \, \cot \beta) \equiv - m_t X_t
\end{eqnarray}
Diagonalising the matrix, we get the mass eigenvalues
\begin{equation}
m^2_{\tilde{t}_{1,2}}
=m^2_t + \frac{1}{2}(M_{\tilde{t}_L}^2+  M_{\tilde{t}_R}^2)+
\frac{1}{4} m^2_Z \cos 2\beta \mp \frac{\Delta}{2}      
\end{equation}
where
\begin{eqnarray}
\Delta^2 &=& \left( M_{\tilde{t}_L}^2 
-  M_{\tilde{t}_R}^2 + \frac{1}{6} \cos 2\beta (8
m^2_W-5m^2_Z) \right)^2 \nonumber \\ && 
  + 4\, m_t^2 |A_t + \mu \cot \beta |^2
\end{eqnarray}
We see a seesaw mechanism here, i.e. one mass is suppressed while the other is enhanced by the off-diagonal terms in the mass matrix. Thus, if the off diagonal terms are large, which is possible because $m_t$ is large, we could get a light stop. 
However, to get a stop which is lighter than other supersymmetric particles, in particular in models with universal soft masses, it is not enough to have large $m_t$ but we need $A_0$ to be also large.

\begin{figure}
\includegraphics[width=0.45\textwidth,height=0.42\textwidth,angle=0]{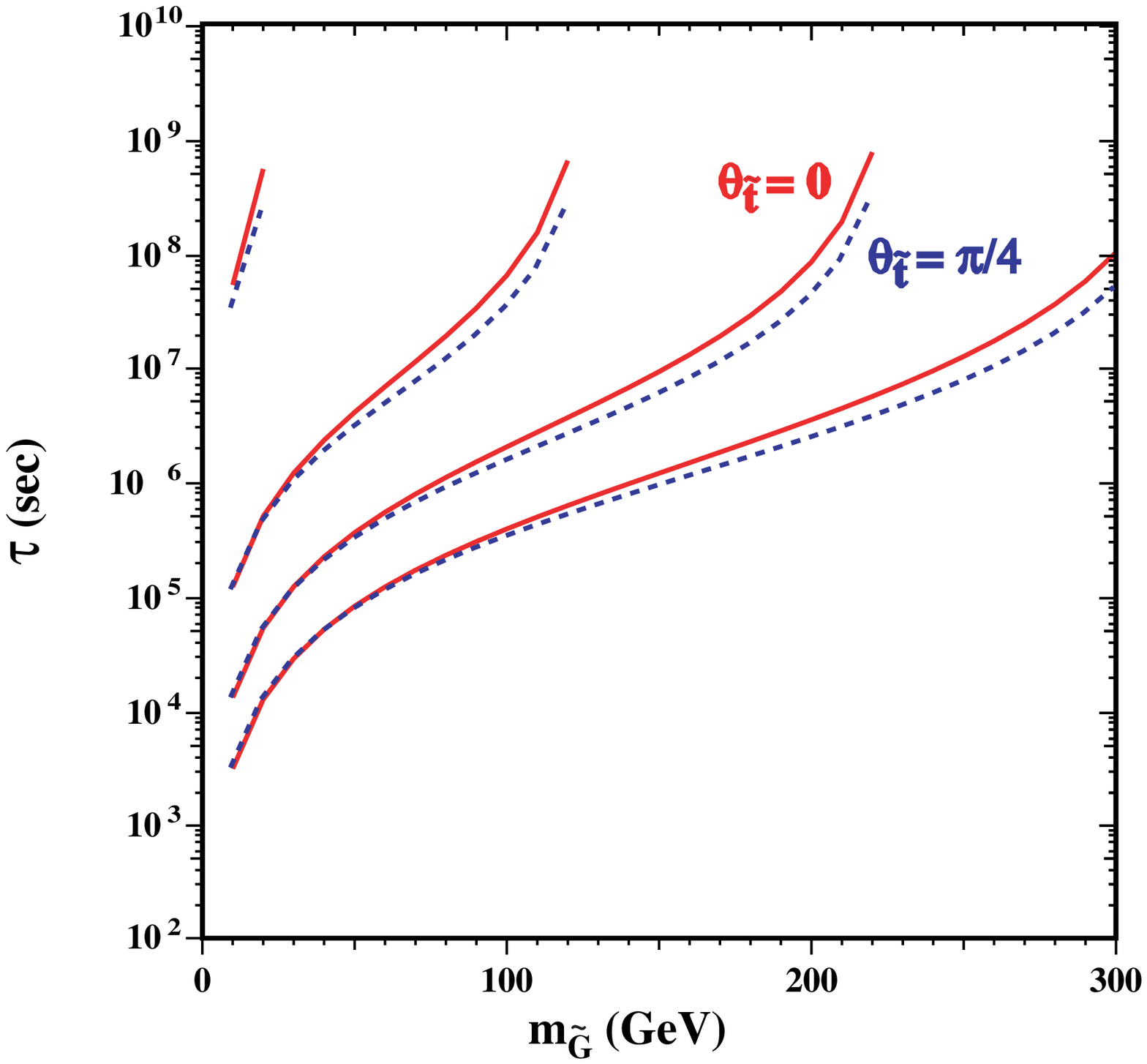}
\vspace{0.2cm}
\includegraphics[width=0.44\textwidth,height=0.42\textwidth,angle=0]{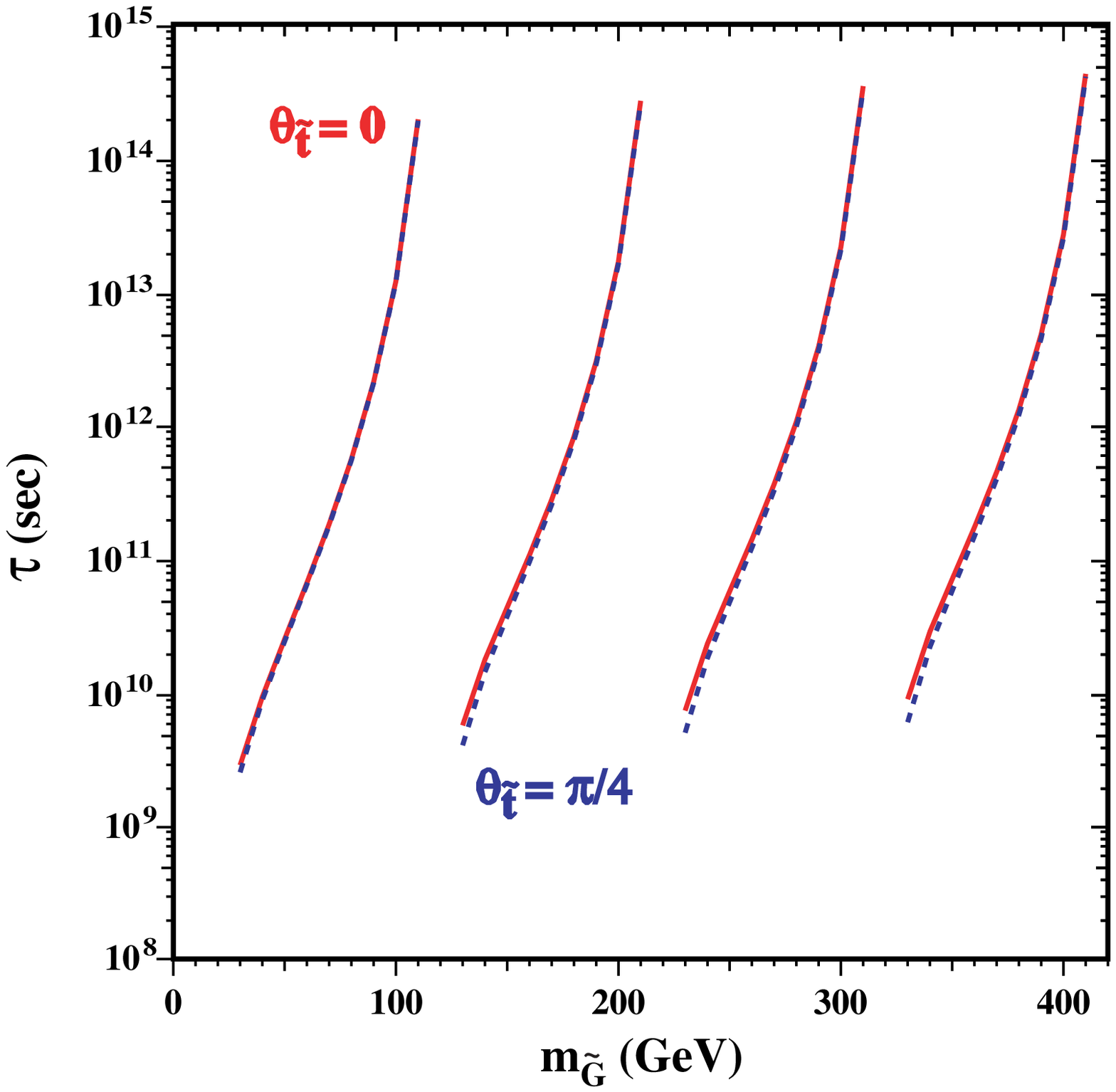}
\caption{Stop lifetime as a function of gravitino mass $m_{\widetilde{G}}$ with fixed values of $\mstop = 200, 300, 400, 500$~GeV respectively from left to right.}
\label{fig:1}       
\end{figure}

The light stop $\tilde{t}_1$ would eventually decay to gravitino $\widetilde{G}$, which is assumed to be the LSP. 
Provided that it is kinematically allowed, the dominant decay mode is the 
2-body decay $\tilde{t}_1 \to \widetilde{G} + t$
\vskip 0.1in
  \begin{center} 
  \begin{picture}(100,50)(0,0)
    \DashLine(0,25)(40,25){3} \Text(5,33)[]{$\tilde{t}_1$}
    \Line(40,25)(80,50) \Text(77,40)[]{$\widetilde{G}$}
    \Photon(40,25)(80,50){1.5}{6}
    \ArrowLine(40,25)(80,0) \Text(78,8)[]{$t$}
  \end{picture}
  \end{center}
with decay rate
\begin{eqnarray}
\Gamma &=& \frac{1}{48 \pi} 
\frac{1}{M_{\rm Pl}^2 m_{\widetilde G}^2 \mstop^3} \nonumber \\
&& \times 
\left[ \left( \mstop^2 - m_{\widetilde G}^2 - m_t^2 \right) 
+ 4 \, \sin \theta_{\tilde t} \, \cos \theta_{\tilde t} \, m_t \, m_{\widetilde G} 
\right]  \nonumber \\
&&\times 
\left[ ( \mstop^2 + m_{\widetilde G}^2 - m_t^2)^2 - 4 \mstop^2 m_{\widetilde G}^2
\right] \nonumber \\
&& \times \left[ ( \mstop^2 + m_t^2 - m_{\widetilde G}^2 )^2 - 4 \mstop^2 m_t^2
\right]^{1/2} 
\end{eqnarray}
If the mass gap $\Delta_m \equiv \mstop - m_{\widetilde G}$ is less than $m_t$ but still greater than $m_W + m_b$, the 3-body decay mode ${\tilde t_1} \to \widetilde{G} + W + b$ becomes the dominant one. It consists of three channels:
\vskip 0.1in
  \begin{center} 
  \begin{picture}(100,50)(0,0)
    \DashLine(0,25)(30,25){3} \Text(5,33)[]{$\tilde{t}_1$}
    \Line(30,25)(80,50) \Text(76,40)[]{$\widetilde{G}$}
    \Photon(30,25)(80,50){1.5}{8}
    \ArrowLine(30,25)(50,15) \Text(40,12)[]{$t$}
    \Photon(50,15)(80,30){1.5}{4} \Text(77,20)[]{$W$}
    \ArrowLine(50,15)(80,0) \Text(66,0)[]{$b$}
  \end{picture}
  \begin{picture}(100,50)(0,0)
    \DashLine(0,25)(30,25){3} \Text(5,33)[]{$\tilde{t}_1$}
    \Photon(30,25)(80,50){1.5}{6} \Text(76,40)[]{$W$}
    \DashLine(30,25)(50,15){2} \Text(40,12)[]{$\tilde{b}_i$}
    \Line(50,15)(80,30) \Text(77,20)[]{$\widetilde{G}$}
    \Photon(50,15)(80,30){1.5}{4}
    \ArrowLine(50,15)(80,0) \Text(66,0)[]{$b$}
  \end{picture}
\vskip 0.2in
  \begin{picture}(100,50)(0,0)
    \DashLine(0,25)(30,25){3} \Text(5,33)[]{$\tilde{t}_1$}
    \ArrowLine(30,25)(80,50) \Text(75,40)[]{$b$}
    \Line(30,25)(50,15) \Text(38,12)[]{$\widetilde{\chi}^+_j$}
    \Line(50,15)(80,30) \Text(77,20)[]{$\widetilde{G}$}
    \Photon(50,15)(80,30){1.5}{4}
    \Photon(50,15)(80,0){1.5}{4} \Text(85,5)[]{$W$}
  \end{picture}
  \end{center}
The 3-body decay rate can be approximated by the following phase space expression 
\begin{equation}
\Gamma_{\rm 3-body} \approx 10^{-23} \, {\rm GeV}^{-6} s^{-1} (\Delta m) \left(
(\Delta m)^2 - m_W^2 \right)^{5/2}
\end{equation}
Note however that for the numerical calculation, we use a more complete analytic formula. 
The numerical results for the stop lifetime are shown in Fig.~\ref{fig:1}.
The first panel shows the lifetime $\tau$ when 2-body decay is dominant. We see that the lifetime can be up to $\sim 10^9$~s. Note that $\tau$ decreases as $\Delta_m$ increases. The figure only shows results with $m_{\widetilde G} \geq 10$~GeV. For smaller gravitino mass we could get lifetime shorter than $10^3$~s. The second panel shows the lifetime through 3-body decay, and we see that the stop can live up to $\sim 10^{15}$~s. 

Of course it is possible that the mass gap between stop and gravitino is even smaller than $m_W$, in which case the 3-body decay channels above are not available. 
In this case we have to consider the 4-body decay ${\tilde t_1} \to {\widetilde G} + b + ({\bar q}q, \ell \nu)$. We can estimate the lifetime using phase space
\begin{equation}
\Gamma_{\rm 4-body} \approx 10^{-30} \, {\rm GeV}^{-8} s^{-1} (\Delta m)^3 \left(
(\Delta m)^2 - m_b^2 \right)^{5/2}
\end{equation}
We shall not forget though, that stop can decay through loop processes as well, such as the following $W$-loop diagram  
  \begin{center} 
  \begin{picture}(100,80)(0,0)
    \DashLine(0,40)(30,40){3} \Text(5,35)[]{$\tilde{t}_1$}
    \DashLine(30,40)(55,60){3} \Text(45,60)[]{$\tilde{q}_{\downarrow}$}
    \Photon(30,40)(55,20){1.5}{6} \Text(42,20)[]{$W$}
    \ArrowLine(55,60)(55,20) \Text(62,40)[]{$q_{\downarrow}$}
    \Line(55,60)(85,75) \Text(80,65)[]{$\widetilde{G}$}
    \Photon(55,60)(85,75){1.5}{6}
    \ArrowLine(55,20)(85,5) \Text(80,13)[]{$c,u$}
  \end{picture}
  \end{center}
However, this process is suppressed by large $m_{\tilde{q}_{\downarrow}}$, especially in the models that we consider, and the small CKM mixing$V_{\rm CKM}$.  So the bottom line is, depending on $\Delta_m$ and the model we use, the stop lifetime could be very long, even longer than the age of the Universes ($O(10^{17}{\rm s})$). 
However, the longer the lifetime the stronger the constraints, such as from the black body spectrum of CMB. 
For simplicity we consider lifetime in the range of $1 - 10^8$~s, and assume that the cosmological constraint is coming only from BBN. 

A long lived stop would hadronize. By taking analogue to heavy quark hadrons we can deduce the lightest hadronic states and approximate their lifetime. The lightest sbaryon is 
$\Lambda_{\widetilde T}^+ \equiv {\tilde t_1} ud$. There are also
$\Sigma_{\widetilde T}^{++,+,0} \equiv \tilde{t}_1 (uu, ud, dd)$ sbaryons which would decay 
strongly into the lightest sbaryons, and the next ones are 
$\Xi_{\widetilde T}^{+,0} \equiv \tilde{t}_1 s (u,d)$ which decay semileptonically
with lifetime $\tau \lappeq 10^{-2}$~s.
There are mesinos, with the lightest one is
${\widetilde T}^0 \equiv \tilde{t}_1 {\bar u}$. We also have 
${\widetilde T}^+ \equiv \tilde{t}_1 {\bar d}$ with lifetime $\tau \simeq 1.2$~s,
and 
${\widetilde T}_s \equiv \tilde{t}_1 {\bar s}$ with lifetime $\tau \simeq 2 \times
10^{-6}$~s. 
Similarly the antistop would hadronize into the corresponding antisbaryons and antimesinos.
So, in about one second after the big bang, only the lightest mesinos ${\widetilde T}^{0(\ast)}$ and lightest sbaryons $\Lambda_{\widetilde T}^\pm$ remain.

\section{Cosmological Implication}
\label{sec:1}

It is well known that a metastable particle can alter the light elements abundances from the standard BBN prediction if it decays during or after the BBN time. The effects can be classified as follows: 
\begin{itemize}
\item
\underline{Photodissociation:}
Electromagnetic showers from the decay can destroy light elements already formed by BBN. The products of this dissociation would then involve in the continuing nucleosynthesis processes.  
\item
\underline{Hadronic showers:}
There could also be effect from quarks and gluons produced by the decay. This effect is of two types: 
\begin{itemize}
\item If the decay happened early, 
the hadronic showers could  change $n/p$ ratio, and we call it hadron injection effect.  
\item If the decay happend around the beginning of the BBN, the hadronic showers could also dissociate light elements, especially $\alpha_{\rm BG}$, and we call this hadrodissociation. 
\end{itemize}
The hadronic showers become relatively unimportant if the decay happens after around $10^6$~s. 
\item
\underline{Catalytic bound state effect:} As pointed out quite recently by Pospelov~\cite{maxim} there could be another important effect if the metastable particle is negatively charged. In this case it can form bound state with
nuclei, hence lowering the Coulomb barrier for certain nucleosynthesis procesess and
introducing photonless final states for radiative capture reactions with much enhanced cross sections.  
\end{itemize}

Now, looking at the stop case, there are two significant facts influencing its BBN effect. First of all, as a strong interacting particle, stop naturally has small relic density due to the large annihilation cross section. 
In turn, the photodissociation and hadronic showers effects would also be relatively small. 
Secondly, as already discussed above, after hadronization only $\Lambda_{\widetilde{T}}^\pm$ and $\widetilde{T}^{0(\ast)}$ left. Because of the mass difference, $\widetilde{T}^0$ and its antiparticle are  
more abundance than $\Lambda_{\widetilde{T}}^\pm$ by $\sim O(10)$ during the early stage.  
The negatively charged $\Lambda_{\widetilde{T}}^-$ is further suppressed by: 
(1) pairing and subsequent annihilation with $\Lambda_{\widetilde{T}}^+$; and 
(2) quark exchange with ordinary hadrons (proton and neutron) into
$\widetilde{T}^0$. 
Note that $\Lambda_{\widetilde{T}}^+$ would also annihilate with antibaryons although not as much as the $\Lambda_{\widetilde{T}}^-$ annihilation with baryons.
Thus at the time of BBN practically only the neutral mesinos around,
and the bound state catalytic effect would become irrelevant. 
The BBN constraint has two parameters: the density of the metastable particle before decay and the lifetime (with fixed baryon density). The stop lifetime depends also on the gravitino mass, which we assume to be a free parameter. As long as the lifetime is not too long (i.e. $\lappeq 10^8$~s) we should be able to satisfy the BBN constraint.

\section{Collider Signatures}

We can also look at the production and detection of metastable stop at colliders. Provided there is enough energy, stop and antistop could be pair produced, and (assuming metastable) they would hadronize before passing the detector.    

There would be neutral as well as charged shadrons (both sbaryons and
mesinos). As the stop shadrons propagate they could undergo  
charge exchange as well as baryon exchange with the background nucleons, for example 
$\widetilde{T} + (p,n) \to (\Lambda_{\tilde{T}} , \Sigma_{\tilde{T}} )
+ n \pi$. We estimate that only about 1/16 of the produced stop-antistop pairs 
yield clear signal of opposite charged pair. 

The charged sbaryon would appear as a charged heavy particle in the muon detector. 
Looking for `slow muon' signal and combine with stop production cross section, one can set the
metastable stop mass lower limit. 
There has been analyses by CDF on the Tevatron Run II data in order to set the mass limit.  
The most recent limit is 250~GeV~\cite{nachtman}, 
although at the time we did our analysis the limit was $m_{\tilde{t}} > 220$~GeV~\cite{phillips}. Both numbers are not published yet, and the limit from the Tevatron Run II could still be changed.

\section{Stop NLSP in Specific Models}

The next obvious question is: can we realize this stop NLSP scenario in a specific supersymmetric model?
In the CMSSM it turns out to be not possible. The reason is because three constraints: stop being the NLSP, the Higgs mass lower bound, and the stop mass lower limit; do not have an overlap allowed region. 

\begin{figure}
\includegraphics[width=0.45\textwidth,height=0.42\textwidth,angle=0]{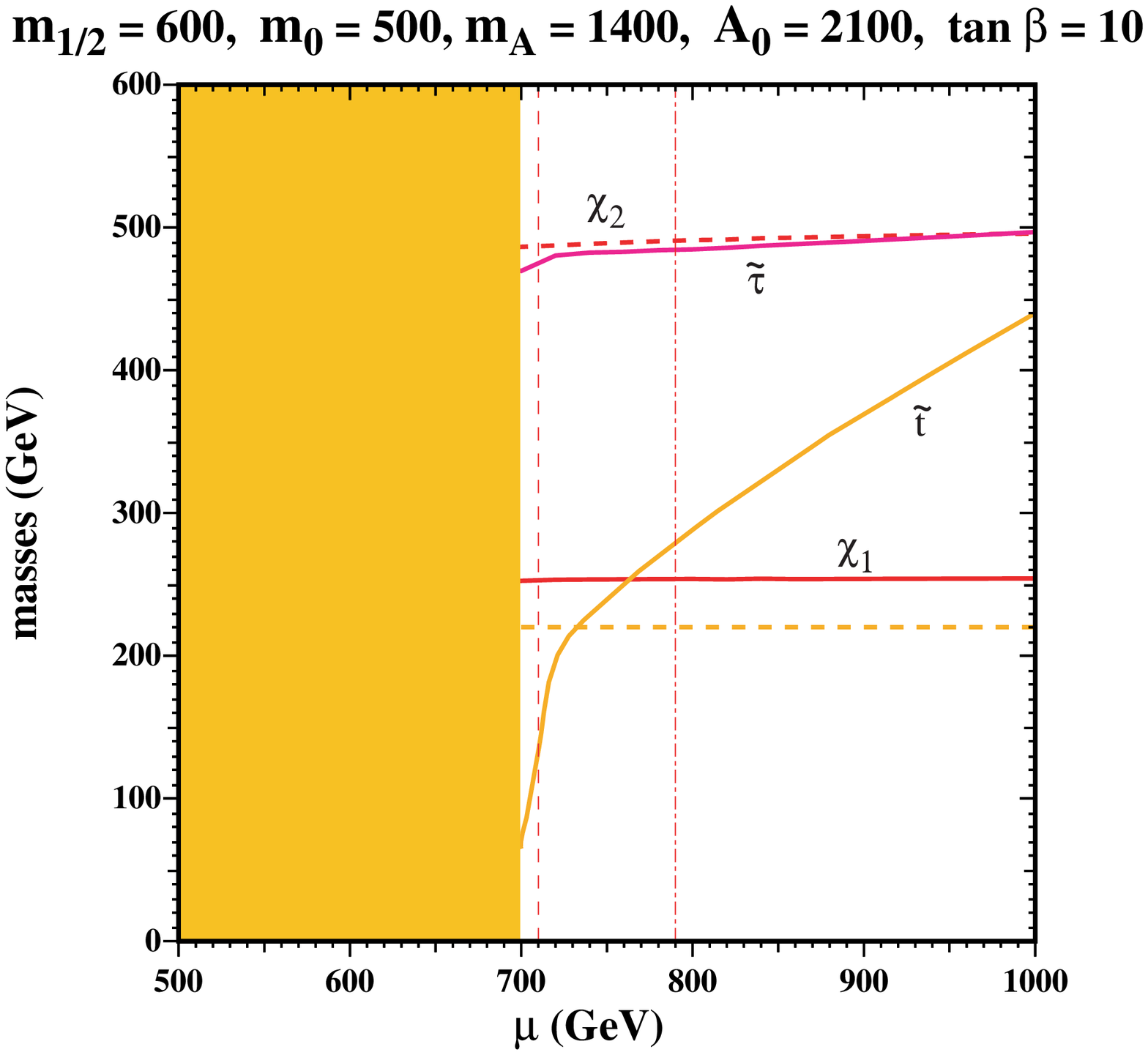}
\caption{NUHM.}
\label{fig:2}       
\end{figure}

In the NUHM model~\cite{nuhm}, which has two more free parameters compared to the CMSSM, we have a better luck and found a small region that is still allowed for $m_{\tilde{t}} > 220$~GeV. It is shown in Fig.~\ref{fig:2}. Note however that this region barely exists if we take 250~GeV as the stop lower mass limit. 

This does not mean that this scenario is in general almost completely excluded. 
Let us take a look at compressed supersymmetry model~\cite{martin}, for example.
This model is motivated by 
SUSY little hierarchy problem, i.e. fine tuning between $|\mu|^2$ and $m_{H_u}^2$ which is required by electroweak symmetry breaking condition. This fine tuning could be aleviated if $m_{H_u}$ is not large. As we know from the RGE, the biggest contribution to $m_{H_u}^2$ comes from $M_3$. 
Thus, this suggests a model with small gluino mass. 
Now, light gluino also leads to light stop, as can be seen through the RGE.
In~\cite{martin}, the author still considered neutralino LSP scenario without gravitino. However, if we choose a bit smaller $M_3$ or a larger but still modest $A_0$, we would be able to get a stop lighter than neutralino. Note however, that the Higgs mass is strongly correlated to the stop sector, so a detail study would be required to reach conclusive statement on this model.   

Perhaps an easier way to satisfy all the constraints is to lift up the neutralino mass instead, i.e. by choosing large bino mass $M_1$ in a nonuniversal gaugino masses model. With large neutralino mass, it would be easy to find a set of parameters with stop lighter than neutralino but still relatively heavy.

\section{Metastable Neutralino}

Looking at Fig.~\ref{fig:2}, we notice that $\chi_1^0 \to \tilde{t}_1 + t$ is not kinematically allowed, hence we could have a long lived neutralino as well. This has implication on cosmology as well as in collider physics. 

In the early universe, neutralino and stop coexist for sometime, with relative density  
$
\Omega_\chi \simeq \frac{1}{3} \ \Omega_{\tilde{t}_1} 
$.
The neutralino could decay directly to gravitino, or to stop which then decay to gravitino - a cascade decay. All of these decays would affect the BBN, and therefore would complicate the analyses.

At colliders, there would be both missing energy signal of neutralino {\it and} slow muon signal of stop shadron. This suggests another benchmark point study for the LHC, with typical mass spectrum shown in Table~\ref{tab:1}.   
%
\begin{table}
\caption{Supersymmetric mass spectrum for an NUHM model with allowed stop NLSP.}
\label{tab:1}       
\begin{center}
\begin{tabular}{ll}
\hline\noalign{\smallskip}
 & {\rm GeV}  \\
\noalign{\smallskip}\hline\noalign{\smallskip}
$M_3$ & 1333  \\
\ldots & \ldots \\
$m_{\chi_1^+}$ & 489 \\
$m_{\chi^0_2}$ & 488 \\
$m_{\tilde{\tau}_1}$ & 482 \\
$m_{\chi^0_1}$ & 253 \\
$m_{\tilde{t}_1}$ & 240 \\
\noalign{\smallskip}\hline
\end{tabular}
\end{center}
\end{table}

\section{Concluding Remarks}

In this review, we have shown that stop NLSP with gravitino dark matter scenario is very
interesting from phenomenological point of view. 
Stop naturally has low relic density due to its
strong interaction, making it possible to satisfy the BBN constraint. 
Metastable stop would hadronize, and the heavier shadron states would decay to the lightest ones. At the time of BBN practically only the
lightest neutral mesino left, hence there is no EM bound state effect.

This scenario is not feasible in the CMSSM because of the
combined constraints from the stop NLSP, the stop mass and the Higgs mass lower bounds.
In the NUHM this scenario is still barely possible. 
Nonuniversal gaugino models with light gluino and/or heavy bino should revive the
feasibility.

%
%

\end{document}